# Penetration depth and effective sample size characterization of UV/Vis radiation into pharmaceutical tablets

R. Brands[1], L. Fuchs[2], J.M. Seyffer[3], N. Bajcinca[4], J. Bartsch[1], U.A. Peuker[3], V. Schmidt[2], M. Thommes[1]

1 Laboratory of Solids Process Engineering, Department of Biochemical and Chemical Engineering, TU Dortmund University, 44227 Dortmund, Germany
2 Institute of Stochastics, Ulm University, 89069, Ulm, Germany
3 Institute of Mechanical Process Engineering and Mineral Processing, Technische Universität Bergakademie Freiberg, 09599 Freiberg, Germany
4 Chair of Mechatronics, University of Kaiserslautern-Landau, Kaiserslautern 67663, Germany
professors.fsv.bci@tu-dortmund.de

**Abstract**

The pharmaceutical industry is moving from off-line quality testing to real-time release testing (RTRT) to improve drug quality while reducing costs. The implementation of RTRT requires advanced in-line process analytics, where UV/Vis spectroscopy has proven its suitability. However, quantification of the sample size requires detailed knowledge of the penetration depth.

In this study, bilayer tablets were produced using a hydraulic tablet press. The lower layer contained titanium dioxide and microcrystalline cellulose (MCC), while the upper layer consisted of MCC, lactose or a combination with theophylline. The thickness of the upper layer was stepwise increased. Spectra from 224 to 820 nm were recorded with an orthogonally aligned UV/Vis probe. Thereby, the experimental penetration depth reached up to 0.4 mm, while the Kubelka-Munk model yielded a theoretical maximum penetration depth of 1.38 mm. Based on these values, the effective sample sizes were determined. Considering a parabolic penetration profile, the maximum volume was 2.01 mm³. The results indicated a wavelength and particle size dependency.

Micro-CT analysis confirmed the even distribution of the API in the tablets proving the sufficiency of the UV/Vis sample size. Consequently, UV/Vis spectroscopy is a reliable alternative for RTRT in tableting.

## 1. Introduction

The pharmaceutical industry is increasingly striving to digitalize its processes [1,2]. This includes autonomous production as well as real-time release testing (RTRT) [3]. Here, process analytical technologies (PAT) are implemented in the process chain to obtain real-time data about the process [4]. Based on these data, a continuous quality assessment is then carried out and finally a release decision is made. This enables to bypass or at least significantly reduce the traditional off-line testing of the drug product, which saves valuable resources and time. [5–7]

Tableting is of particular relevance within the scientific scope of pharmaceutical digitalization, as the tablet constitutes one of the most prevalent oral solid dosage forms [8]. The implementation of PAT to monitor critical quality attributes (CQAs) for RTRT through spectroscopic methods is of notable importance [9,10]. These enable non-invasive, non-destructive, and comparatively fast measurements [11]. In this context, the in-line measurement of CQAs such as active pharmaceutical ingredient (API) content, porosity and hardness of the tablet is crucial to enable a quality assessment of the final drug product during the production process.

Through the implementation of near-infrared (NIR) spectroscopy technology, the API content of tablet formulations with several excipients can be monitored [12,13] as well as physical CQAs such as porosity and hardness [14–17]. However, multivariate analysis in the form of partial least squares regression is necessary in this context. With reflectance NIR microscopy, the depth from which the spectral information is obtained, referred to as penetration depth, ranges up to 777 µm [18]. The penetration depth of NIR radiation into wheat flour was characterized and a wavelength dependency was identified. Here, the values ranged from 0.5 to 2.5 mm [19]. Furthermore milk powder was analyzed and characterized with similar penetration depths [20]. Likewise, a wavelength and API content dependency was observed for tablets with API, whereby the measurements were performed in both transmission and reflection mode, resulting in higher penetration depth values for transmission measurements [21]. Further investigations indicated that 90% of the spectral information in tablets results from depths lower than 0.25 mm [22,23].

However, a promising alternative to NIR spectroscopy is UV/Vis spectroscopy due to univariate data evaluation and short exposure times in the range of milliseconds [24]. Therefore, UV/Vis spectroscopy can be utilized to determine the API content in real time via diffuse reflection in tableting. Here, the reflectance in the UV range is evaluated univariatly and correlated with the API content [25]. In addition, physical CQAs such as porosity and hardness can be monitored in-line after converting the Vis spectrum into the CIELAB color space [26].

However, in the case of UV/Vis spectroscopy, the penetration depth in pharmaceutical materials or tablets is so far unknown. Nevertheless, this information is crucial in order to define the effective sample size and thus state how many molecules are included in a sample. The experimental penetration depth should be distinguished from the theoretically achievable penetration depth of radiation into materials, disregarding measurement uncertainties and technical resolution limitations. This depth can be determined by applying the Kubelka-Munk model [27], which considers the influence of absorption and scattering for the calculation of the reflectance [19,28].

The aim of this study was to characterize the penetration depth of UV/Vis radiation in tablets and thereby to determine the effective sample size. For this purpose, bilayer tablets were produced with increasing thickness of the upper layer and its composition was varied. Objectives of these investigations were determining the influences of the particle size of the formulations, of the deformation behavior and of the addition of UV absorbing substances on the penetration depth. Furthermore, the present study investigates the spatial distribution of the API within the tablet. For this purpose, micro-CT images were utilized to correlate the penetration depth with the CQAs measured by UV/Vis spectroscopy.

## 2. Materials and methods
### 2.1. Materials and material characterization

The bilayer tablets consist of a lower layer and a corresponding upper layer. Here, the lower layer invariably consists of 80 wt% titanium dioxide (1171, Kronos, Leverkusen, Germany) and 20 wt% microcrystalline cellulose (Emcocel 90M, JRS Pharma, Rosenberg, Germany). The upper layers are obtained from different compositions. Microcrystalline cellulose (Emcocel 90M, JRS Pharma, Rosenberg, Germany) with three different particle sizes was investigated. For particle size fractioning, 90 and 180 µm sieves were used. The unfractionated microcrytalline cellulose is referred to as MCC, the coarse fraction as MCC coarse and the fine fraction as MCC fine. α-Lactose monohydrate was utilized as a comparatively brittle material

(FlowLac 90, Meggle, Wasserburg am Inn, Germany) in combination with 0.5 wt% magnesium stearate (Ligamed MF-2-V, Peter Greven, Bad Münstereifel, Germany) as a lubricant. In addition, two formulations consisting of either 89.5 wt% lactose or microcrystalline cellulose containing 10 wt% theophylline monohydrate (Theophylline Monohydrate, BASF, Ludwigshafen, Germany) as a radiation-absorbing substance were tested. The blend of lactose and theophylline is referred to as LT and that of MCC and theophylline as MT.

The characterization of the formulations with respect to the particle size distribution was carried out using the Mastersizer 3000 (Malvern Panalytical, Malvern, UK). For dispersion purposes, the dry dispersing unit Aero S (Malvern Panalytical, Malvern, UK) was utilized for microcrystalline cellulose, α-lactose monohydrate, magnesium stearate and theophylline monohydrate and the wet dispersing unit Hydro EV (Malvern Panalytical, Malvern, UK) was used for titanium dioxide. Each powder component was analyzed in quadruplicate, while measurements were performed according to the monograph of the European Pharmacopoeia [29].

### 2.2. Tableting

Tablets were produced on a manual hydraulic tablet press (PW, Paul Weber Maschinen und Apparatebau, Remshalden, Germany). The punches had a 13 mm diameter flat compression surface. Initially, 800 mg of the lower layer formulation was filled in the die and the lower layer was pre-compressed with a compression force of 5 kN and a dwell time of one minute. Subsequently, the corresponding quantity for the upper layer was added to the pre-compressed lower layer. For upper layers containing microcrystalline cellulose, masses of 10 to 40 mg were tested. In the case of formulations containing lactose, the range from 15 to 50 mg was investigated in equidistant 5 mg steps. The different masses result from the bulk densities of microcrystalline cellulose and lactose and the resulting heights and mechanical strength of the upper layers. Finally, the upper layers were pressed directly onto the lower layer with a main compression force of 18 kN and a dwell time of one minute. For each test series, four tablets per mass level were produced, resulting in a total of 32 tablets per test series. In addition, tablets consisting only of the upper or lower layer were produced and thereafter used as a reference.

After the production and the following UV/Vis measurement, the tablets were characterized regarding their weight with a scale (MCE224S, Sartorius Lab Instruments, Göttingen, Germany) and thickness with a gauge (ABS AOS, Mitutoyo, Kawasaki, Japan). Then, the layers were separated manually. Thereby, the thickness of the upper layer could be determined individually as well as the porosity in accordance with the European Pharmacopoeia and with respect to the true density characterized via helium pycnometry according to European Pharmacopoeia 11 (Chapter 2.9.23).

### 2.3. UV/Vis spectroscopy

All spectra were measured directly after production in the die to ensure reproducible positioning under the probe. For this purpose, a corresponding sample holder was used. With a micrometer screw, the distance of 4 mm between the probe and tablet was adjusted and kept constant. The UV/Vis probe (Inspectro X, ColVisTec, Berlin, Germany) is implemented orthogonally to the upper tablet layer surface. The measurement duration was one minute with a sampling frequency of 2 Hz and an exposure time per measurement of 70 ms. Prior to the measurement, the UV/Vis probe was calibrated with a black $R_{black}$ and white $R_{white}$ tile (ColVisTec, Berlin, Germany) in order to determine the relative diffuse reflectance %R (Eq. 1).

| $$\%R = \frac{R - R_{black}}{R_{white} - R_{black}} * 100\%$$ | 1 |
|---|---|

Pre-processing of the spectra was performed to reduce the influence of scattering effects on the spectra and thus highlight the chemical information. For this purpose, the spectra were subjected to standard normal variate correction (Eq. 2) [30].

$$f(\%R) = \frac{(\%R_\lambda - \%\bar{R})}{\sqrt{\frac{\sum_{\lambda=1}^{n}(\%R_\lambda - \%\bar{R})^2}{n-1}}} \quad \quad 2$$

Here, the spectra are centered using the mean value and the standard deviation is used for scaling. In this context, $\%R_\lambda$ represents the relative diffuse reflectance at a certain wavelength λ, $\%\bar{R}$ the average relative diffuse reflectance and $n$ the number of wavelengths. Figure 1 depicts the experimental setup for investigating light penetration in a tablet, where an incident light beam with intensity $I_0$ strikes the sample surface perpendicularly. Within the tablet, the light is scattered and absorbed, with intensity i(x) representing the forward-propagating light and j(x) describing the backscattered light at depth x. The reflectance $R_g$ corresponds to the reflectance of the lower layer.

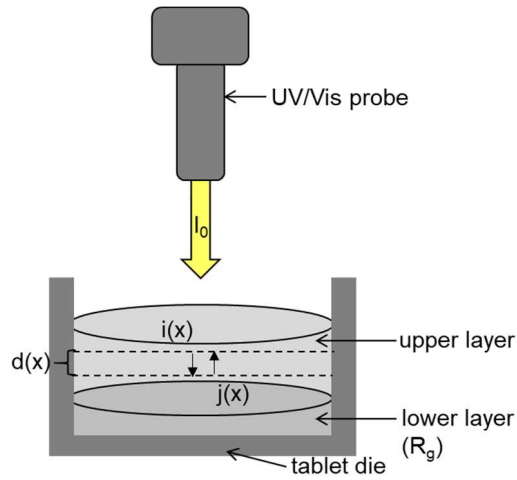

Figure 1: UV/Vis probe setup for bilayer tablet characterization.

### 2.4. Kubelka-Munk Theory

The Kubelka-Munk model, developed for a monochromatic case to investigate the influence of paints on the color of a background, is applied in the following. In this model, an upper layer of thickness z is exposed to a radiation intensity $I_0$ (the incident light intensity). Considering now an infinitesimal layer dx, two opposing light fluxes i and j exist, whereby the light flux j is created by reflection. As light interacts with the material, part of the incident light is absorbed by the material, while another part is scattered and reflected. The changes in the light fluxes i and j over the layer thickness can be described using these absorption coefficient k and scattering coefficient s. Consequently, the change in radiation intensity when passing through an infinitesimal layer is composed of absorption, scattering, and the light flux scattered in the opposite direction. These relationships lead to the following system of differential equations at a depth in the layer x with the corresponding coefficients k and s (Eq. 3 and 4) [31].

| | |
|---|---|
| $-di(x) = -(s+k)i(x)dx + sj(x)dx$ | 3 |
| $dj(x) = -(s+k)j(x)dx + si(x)dx$ | 4 |

The reflectance R of the sample is obtained by explicitly solving the system of differential equations under the assumption that the layer thickness of the sample is considered finite and the reflectance of the background $R_g$, in this case the lower layer, affects the diffuse reflection (Eq. 5-7) [19,28].

| $$R = \frac{1 - R_g(a - b\,coth(bsy))}{a - R_g + b\,coth(bsy)} \text{ with}$$ | 5 |
|---|---|
| $$a = \frac{k+s}{s}$$ | 6 |
| $$b = \sqrt{a^2 - 1}$$ | 7 |

The reflectance of the background $R_g$ plays a significant role in determining the overall reflectance R of the system, especially when the layer thickness y is small. As the sample thickness increases, the contribution from the background decreases until it no longer influences the measured spectra, leading to the concept of penetration depth.

The theoretical penetration depth describes how deep light penetrates into a material, based on the absorption and scattering coefficients, regardless of whether this depth contributes to the measurement. In this context, the Kubelka-Munk model allows for calculating this reflectance at infinite thickness $R_\infty$, where the influence of the lower layer is negligible. The experimental penetration depth, on the other hand, refers to the actual depth from which useful information still contributes to the measurement signal

### 2.5. Experimental penetration depth

According to the Kubelka-Munk model, the information contribution of the lower layer decreases as the thickness of the upper layer increases. Once the experimental penetration depth has been reached, the spectrum corresponds to that of the pure upper layer. For this purpose, spectra of pure upper layers with a height of 2 cm were recorded to ensure that the layer thickness corresponds to that of an infinite thick layer without containing any information from the material underneath. An appropriate mean value spectrum was generated from these spectra of the pure upper layers including standard deviation. The corresponding reflectance range of this pure upper layer is considered as the threshold in the following (Fig.2, grey area). Subsequently, the tablets were produced with increasing top layer thicknesses and the spectra were recorded. Afterwards, the reflectance values for each wavelength from 224 to 820 nm were plotted against the thickness of the upper layer. The experimental penetration depth is exceeded when the reflectance values of the upper layers reach the reflectance range of the pure upper layer, as no further differentiation can be performed in this case (Fig. 2, 1). Therefore, this penetration depth is equivalent to the thickness of the upper layer, given that a distinction can still be made between the reflectance value of the pure upper layer and the bilayer tablet (Fig. 2, 2). The penetration depth was determined for each wavelength from 224 to 820 nm (Fig. 2, 3). For its representation over the wavelength, a moving average was calculated over 10 wavelengths.

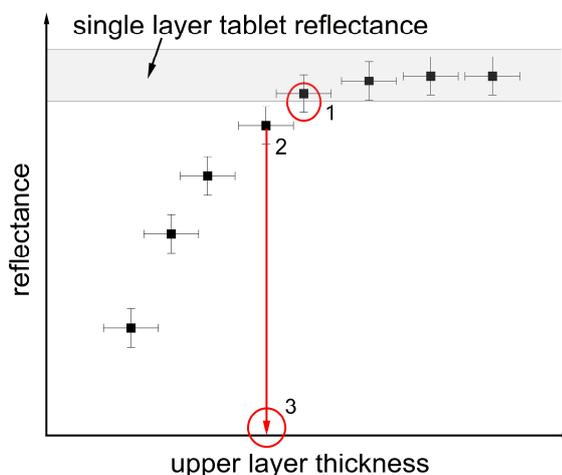

Figure 2. Schematic illustration of the experimental information for the depth characterization.

### 2.6. Theoretical penetration depth

The Kubelka-Munk model (Eq. 5) was fitted to the individual data points applying the least squares method using Origin 2024 software (OriginLAB, Northampton, MA, USA) to determine the theoretical penetration depth, when the reflectance values reaches $R_\infty$. This procedure was conducted for three selected wavelengths, namely 331, 587 and 815 nm. Thereby, the values for k, s and $R_g$ were determined. In addition, $R_\infty$ was calculated according to equation 8 [19,28]. Here, the upper layer is assumed to be thick enough, so that the lower layer has no effect on the spectrum. Thus, the final $R_\infty$ only depends on k and s.

$$R_\infty = 1 + \frac{k}{s} - \sqrt{\frac{k^2}{s^2} + 2\frac{k}{s}} \qquad 8$$

The corresponding theoretical penetration depth was calculated using a limit value formation. This was carried out numerically, whereby y was increased by 0.01 mm in each step. Here, it was defined that the final R is reached as soon as the difference between R and $R_\infty$ is less than $10^{-6}$.

### 2.7. Effective sample size characterization

The effective sample size $V_{effective}$ was calculated for the experimentally determined penetration depths. The distance from the probe to the tablet was defined to be 4 mm and the light cone was described with an angle of 60°. The penetration of the light beam into the sample was assumed to be parabolic, resulting in a lower penetration depth at the edge compared to the center. $V_{effective}$ was thus determined by integrating the intensity profile I(z) along the information depth y. Hereby x represents the depth in the tablet, while x=0 mm equals the tablet surface and y the maximum penetration depth.

$$V_{effective} = A \int_0^y (1 - \frac{x}{y})^2 dx \qquad 9$$

Furthermore, a theoretical investigation was carried out for tablets with a diameter of 8 mm and a height of 4 mm for the MT and LT formulations containing the API theophylline monohydrate. In this theoretical investigation, it was evaluated how many of the total API molecules in the tablet are included in the effective sample size by UV/Vis spectroscopy.

### 2.8. Spatial API distribution

In order to utilize information derived from UV/Vis spectroscopy on tablet surface composition effectively, an analysis of the tablet's microstructure in relation to its spatial position is crucial. This analysis was conducted on the LT tablet formulation using micro-computed tomography (micro-CT) imaging, followed by segmentation into API and excipients using a 3D U-Net model [32]. This segmentation enabled the spatial distribution analysis of tablet components. The data for this was acquired by fixing a tablet of the LT formulation within a low-attenuation polymeric sample holder tube using foam rubber padding.

Three-dimensional micro-CT imaging was done using a Zeiss Xradia Versa 510 (Carl Zeiss Microscopy GmbH, Oberkochen, Germany) equipped with an optical microscope unit and multiple magnification lenses. A 4x lens was selected, providing a zoomed-in field of view to achieve a higher resolution scan. Thus, the setup results in interior tomography or region-of-interest (ROI) tomography, as only a portion of the tablet is captured in each scan.

Covering the full width of the tablet requires five individual scans with 15-20% overlap along the z-axis (horizontally in Figure 1), which were subsequently vertically stitched. The parameters for each scan were as follows: acceleration voltage of 80 kV at 7 W, with a Zeiss standard low-energy filter (LE2) applied. Each individual scan includes 1601 projection images over the range of 360°, acquired with an exposure time of 5 seconds per image. The acquired data was reconstructed using a filtered back-projection algorithm within the Zeiss Reconstructer software. Automatic stitching of the scans is performed using the same software, resulting in an isotropic voxel size of 2.11 µm and a measured volume of about 2 x 2 x 7 mm.

Once image data acquisition and reconstruction were complete, the 3D image was segmented using a 3D U-Net deep learning model. The model was trained on partially hand labeled 2D slices of the CT-image in order to distinguish between the API and excipients in the 3D image.

Preprocessing of the CT-image employs the use of a non-local means filter [33]. The hand labeled data were then used to run 2D slice-based model training by minimizing a focal loss [34] with gradient descent. See [35] for a detailed description of the network training procedure.

## 3. Results and discussion
### 3.1. Particle size distribution

The particle size distributions were evaluated for the respective raw materials (tab.1). Here, unfractionated microcrystalline cellulose was found to have a $d_{50}$ of 122.5 µm. Sieve fractionation resulted in a coarse fraction with a $d_{50}$ of 159.25 µm and a fine fraction with a $d_{50}$ of 68.50 µm. The span of the distributions of these MCC powder samples are in a comparable range of 1.62 to 1.80. The used lactose quality is characterized by a $d_{50}$ of 133.75 µm, thus in a comparable size range as the unfractionated MCC. However, the distribution of lactose is narrower compared to that of MCC, characterized by a span of 1.07. For the model API theophylline monohydrate used in this study, a $d_{50}$ of 99.90 µm and comparably larger particles in the coarse fraction with a $d_{90}$ of 381.33 µm were found. This results in a broad distribution with a span of 3.48. Titanium dioxide as a white pigment exhibits the smallest particles in this study with a $d_{50}$ of 0.60 µm. In contrast, the lubricant magnesium stearate is characterized by a $d_{50}$ of 6.87 µm and the broadest distribution.

Table 1: Particle size distributions of used raw materials. n=4; $\bar{x}$±s

|  | $d_{10}$ [µm] | $d_{50}$ [µm] | $d_{90}$ [µm] | span [-] |
|---|---|---|---|---|
| **Microcrystalline cellulose** | 36.30 ± 0.25 | 122.50 ± 0.73 | 256.75 ± 0.68 | 1.80 ± 0.01 |
| **Microcrystalline cellulose coarse** | 49.08 ± 1.19 | 159.25 ± 2.22 | 307.25 ± 4.50 | 1.62 ± 0.01 |
| **Microcrystalline cellulose fine** | 25.28 ± 0.53 | 68.50 ± 0.75 | 142.75 ± 1.26 | 1.72 ± 0.01 |
| **Lactose monohydrate** | 71.83 ± 2.15 | 133.75 ± 3.30 | 215.50 ± 6.35 | 1.07 ± 0.01 |
| **Theophylline monohydrate** | 33.67 ± 1.33 | 99.90 ± 4.43 | 381.33 ± 22.23 | 3.48 ± 0.05 |
| **Titanium dioxide** | 0.29 ± 0.01 | 0.60 ± 0.01 | 1.02 ± 0.01 | 1.22 ± 0.01 |
| **Magnesium stearate** | 2.12 ± 0.02 | 6.87 ± 0.12 | 34.60 ± 2.43 | 4.72 ± 0.24 |

### 3.2. Raw material spectra

All raw materials were compressed into tablets and the spectra were measured accordingly. The corresponding spectra after SNV transformation are presented in Figure 3. For all raw materials, peaks are visible in the UV range, whereas baselines are visible in the Vis range. The three different MCC qualities and the lactose utilized exhibit similar spectral profiles, characterized by a baseline in the Vis range and a peak in the UV range at approximately 270 nm. However, these substances do not feature any pronounced chromophoric systems.

Therefore, the peak in the UV range is due to the relative measurement in relation to the white tile, which absorbs more UV radiation compared to the tablets. The model API theophylline monohydrate exhibits a peak in the UV range and two minima followed by baseline in the Vis range. The lower layer consists of 80 wt% titanium dioxide, which is characterized by strong scattering properties in the Vis range. Accordingly, a comparably stronger reflection is recognizable in the Vis spectrum with a baseline shift towards high reflectance values. Furthermore, a minimum in the UV range at approximately 345 nm is recognizable. Hence, the lower layer can be distinguished from the upper layer in terms of raw material spectra. Consequently, the influence of the upper layer thickness and thus the information fraction of the lower layer on the spectrum should be recognizable. The regions at 331, 587 and 815 nm are particularly suitable for further evaluation.

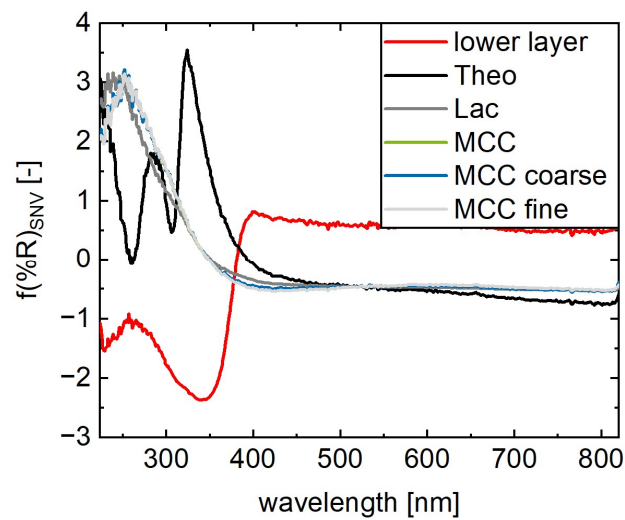

Figure 3: Raw materials spectra consisting of relative diffuse reflectance over the wavelengths.

### 3.3. Tablet characterization

The upper layers weighed 10 to 40 mg for the lactose containing formulations and 15 to 50 mg for the MCC formulations. This resulted in layer thicknesses of approximately 0.12 to 0.4 mm. The porosities of the formulations containing MCC were set to approximately 50%, whereas a porosity of approximately 40% was set for the formulations containing lactose. Tablet height and porosity are a product of the compaction behavior of the powder components. The objective was to produce mechanically stable layers despite thin layer thicknesses.

### 3.4. Experimental penetration depth evaluation

The penetration depth was determined experimentally by producing bilayer tablets. Initially, a tablet consisting only of the lower layer was produced and measured followed by increasing the thickness of the upper layer in discrete steps by adding more powder formulation. The spectrum of the bilayer tablet was then measured accordingly. Figure 4 illustrates the changes in the spectrum caused by increasing the upper layer thickness of the MCC coarse formulation exemplary. First, the spectrum consists only of information from the lower layer. However, the spectrum changes when an upper layer is pressed on top of the lower layer and thus the spectrum is now a superposition of spectral information from both layers. By further increasing the upper layer thickness, the spectrum becomes increasingly similar to the spectrum of the pure MCC coarse tablet. Consequently, the amount of information of the lower layer decreases until finally the upper layer reaches a thickness where no more information of the lower layer is in the spectrum and it is no longer distinguishable between the spectra of the tablets.

Thereby, the reflectance values below 380 nm increases with increasing thickness, while above this threshold a reversed correlation is recognizable.

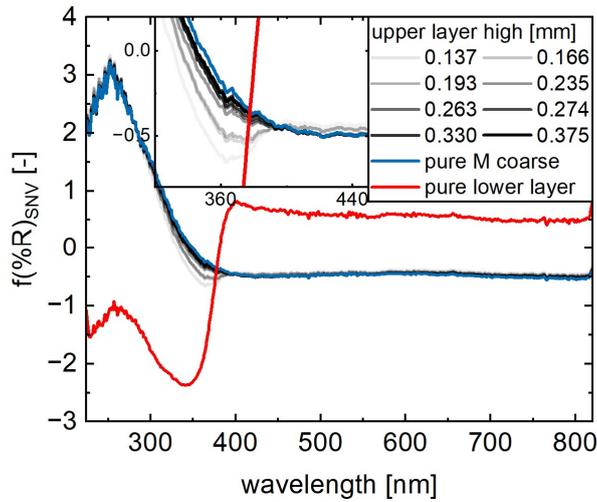

Figure 4: Diffuse reflectance decrease with increasing upper layer high.

In order to determine the penetration depth, the reflectance value per wavelength was evaluated. Figure 5 shows exemplary the determination of the penetration depth for the wavelengths 331 and 587 nm. Nevertheless, the evaluation procedure described here was performed for all wavelengths from 224 to 820 nm. The gray region in the diagrams represents the mean reflectance value of the pure tablets consisting only of the upper layer plus minus the standard deviation. It can be observed that the reflectance value at a wavelength of 331 nm increases with the increasing thickness of the upper layer. Furthermore, the values converge towards a value that is located in the gray area. Here, it is no longer possible to differentiate between the pure upper layer and the bilayer tablet. Consequently, no information from the lower layer is recognizable in the spectrum. The penetration depth is accordingly at the next smaller thickness of the upper layer, where the standard deviation does not reach the gray area. For the wavelength 331 nm, the penetration depth is 0.26 mm. In the case of wavelengths above 380 nm, such as 587 nm (fig.5, right), an inverse correlation between reflectance and upper layer thickness is recognizable. Here, the penetration depth is 0.19 mm.

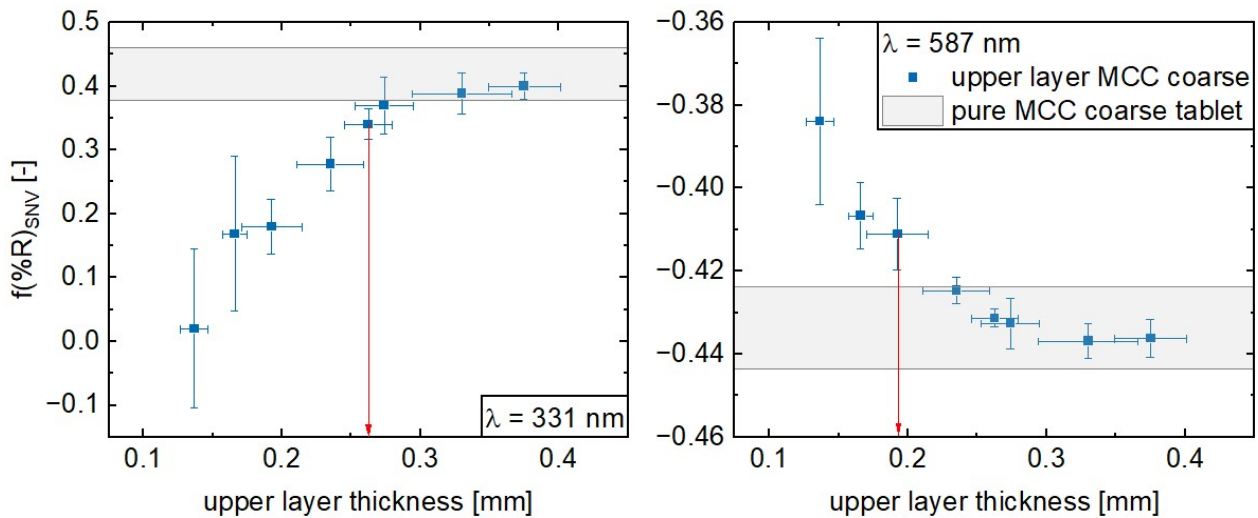

Figure 5: Experimental penetration depth characterization for 331 nm and 587 nm. n=4; $\bar{x} \pm s$

This procedure was applied for all wavelengths of all formulations. The results are illustrated in Figure 6. Note that for mechanical reasons, layers thinner than 0.1 mm could not be produced and therefore no penetration depths were calculated. Also, no penetration depths around 380 nm could be calculated as the spectrum of the lower layer and the corresponding upper layer are equal here. However, the precise location of this range depends on the raw materials used.

Figure 6 illustrates the wavelength dependency of the penetration depth due to the interaction of absorption, transmission and scattering with the material. In the UV range, the penetration depth is lower compared to higher wavelengths due to the strong absorption of radiation by molecular transitions in the material and strong scattering, which limits the penetration of the UV radiation. However, for wavelengths in the transition region between UV and Vis, absorption decreases and scattering becomes less pronounced. This may lead to a gradual increase in penetration depth. In Figure 6, however, the value of the penetration depth appears to decrease, which could have been artificially caused by the previously described similarity of the spectra. The Vis radiation interacts less effectively with the particles, allowing higher transmission in certain areas. Nevertheless, between 500 and 700 nm, the penetration depth remains comparatively low due to increased incoherent scattering, which affects the light transmission. During incoherent scattering, radiation is scattered in all directions. In the NIR region, the penetration depth increases since scattering and absorption are decreased at these longer wavelengths. Coherent scattering is more important here, where the radiation is scattered but not fully redirected from the original path through the material, which leads to a comparable increase in penetration depth. Above 800 nm, the penetration depth increases further, indicating a minimal interaction between the radiation and the particles, allowing deeper light transmission and higher penetration depths. Consequently, the interaction of UV, Vis and NIR radiation with the tablet differs and influences the penetration depth.

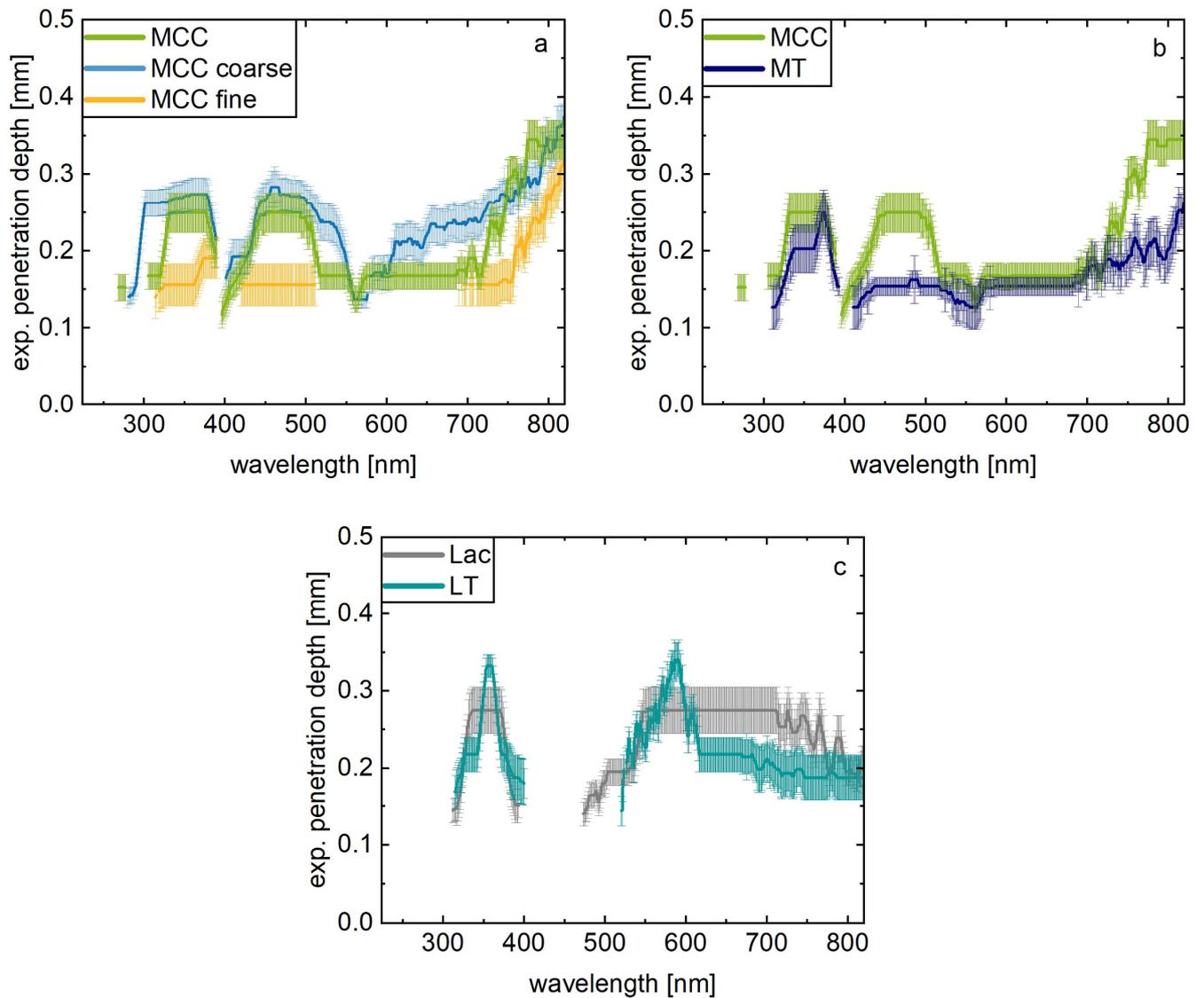

Figure 6: Experimental penetration depth results for all tested materials including moving average over 10 wavelengths. $\bar{x}$±s

The influence of the MCC particle size can be observed in figure 6a, where the penetration depth for all three MCC qualities is overall between 0.1 and 0.4 mm. The characteristics in the UV, Vis and NIR regions appear to be similar, however, the value for MCC coarse is shifted towards higher and for MCC fine towards lower penetration depths. This may be caused by the interaction of radiation with the particles. Smaller particles lead to increased omnidirectional scattering of radiation, thereby reducing the penetration depth of the radiation in the tablet, as backward reflection also occurs. In addition, we assume that this scattering behavior is reinforced by the presence of smaller but more prevalent air voids. Larger particles, on the other hand, lead to increased forward scattering and therefore to an increased penetration depth. These findings are consistent with the general scattering principles of the Kubelka-Munk model.

The addition of 10 wt% theophylline leads to a decreased penetration depth compared to the formulation without API (fig. 6b), particularly below 300 nm. Here, the value remains under the limit for mechanically stable tablets. In this particular region, theophylline absorbs radiation, resulting in a reduced penetration depth. Additionally, theophylline has a comparatively smaller particle size, which leads to increased scattering.

Lactose differs from MCC in terms of its deformation behavior, among other things. Lactose can primarily be described as brittle while MCC is connected to a plastic deformation behavior. In this study, the particle sizes are in comparable ranges. Lactose and MCC show similar patterns in the UV range when comparing figure 6 a and b. However, a higher penetration depth is recognizable in the Vis range. The inverse course of the pattern for Lactose in comparison to MCC thereby indicates a reduction of the scattering effects. Furthermore, a decreases in penetration depth with increasing wavelength towards the NIR range is observable. The pattern after the addition of theophylline remains similar to that of lactose.

Overall, the penetration depths for all tested formulations were between 0.1 and 0.4 mm. Thereby, an influence of the particle size as well as the wavelength can be observed. Thus, surface tablet attributes instead of global attributes are measured.

### 3.5. Theoretical penetration depth

The theoretical penetration depth was determined according to the Kubelka-Munk model, as the experimental determination is naturally subject to measurement limitations and uncertainties. Therefore, the model equation (Eq. 5) was fitted to the experimental data and the corresponding coefficients k and s were determined (Fig. 7). Subsequently, the reflectance at infinite layer thickness $R_\infty$ was determined according to equation 8. The theoretical penetration depth at this reflectance was then determined numerically. This approach was carried out exemplarily for the three wavelengths 331, 587 and 815 nm.

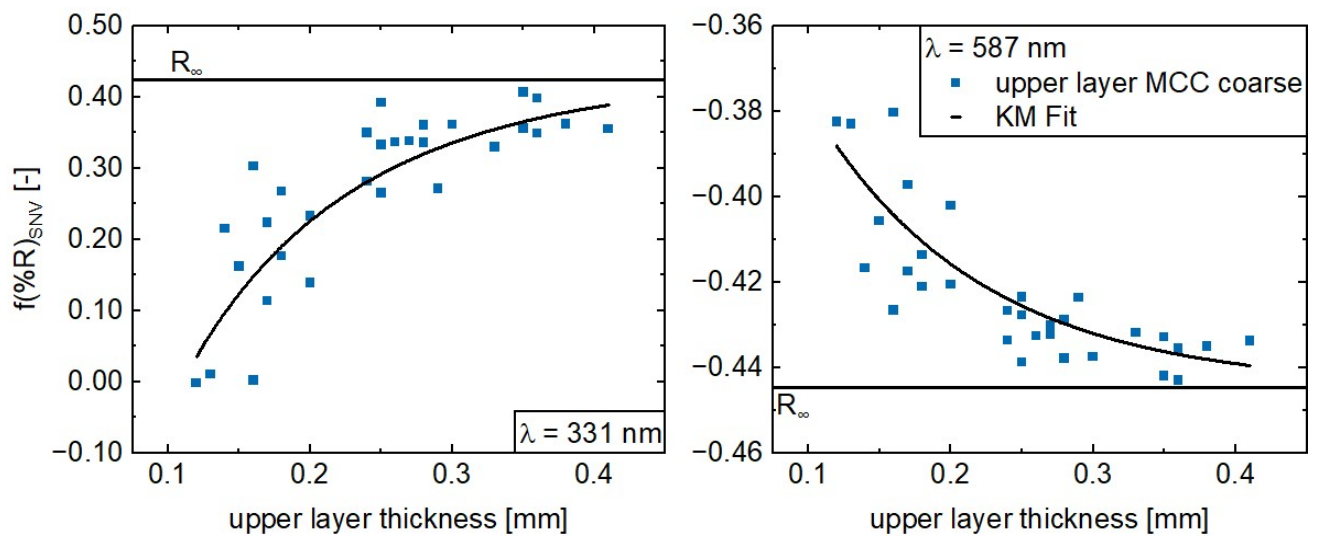

Figure 7: Theoretical penetration depth characterization exemplary for the MCC coarse formulation and 331 nm (left) and 587 nm (right) using the Kubelka-Munk (KM) model.

Table 2: Theoretical penetration depths evaluated using the Kubelka-Munk model for 331, 587 and 815 nm.

|  | theoretical penetration depth y($R_\infty$) mm | | |
| --- | --- | --- | --- |
| Formulation | 331 nm | 587 nm | 815 nm |
| MCC | 0.69 | 0.34 | 0.33 |
| MCC coarse | 1.26 | 1.06 | 1.38 |
| MCC fine | 0.68 | 0.60 | 1.06 |
| MT | 0.73 | 0.53 | 0.72 |
| Lac | 0.78 | 0.65 | 0.57 |
| LT | / | / | 0.79 |

The theoretical penetration depths were found to be between 0.25 and 1.38 mm, as observed in Table 2. In particular, MCC coarse demonstrated the highest penetration depth with 1.38 mm, while MCC fine exhibited the lowest. The previously described effect of particle size on scattering also influenced the observed decrease in penetration depth with decreasing particle size. According to Kubelka-Munk, these results are to be expected, as the penetration of radiation into material is dependent on the scattering behavior. In case of the LT formulation, no theoretical penetration depth could be determined at 331 and 587 nm due to model fitting issues.

In general, the theoretical penetration depth is higher compared to the experimental, considering that this is a theoretical investigation, where no measurement limitations are considered.

### 3.6. Effective sample size

The effective sample size describes the proportion of the tablet that contributes to the spectral information at a defined penetration depth and varies from 0.84 to 2.01 mm$^3$ (tab.3). Here, the experimental penetration depth and the parabolic radiation penetration pattern are taken into account. Thereby, the sample size increases with the penetration depth. For the formulations MT and LT, the percentage of measured theophylline molecules can be determined for a model tablet with a diameter of 8 mm and a height of 4 mm. In the case of the MT formulation, 0.5, 0.42 and 0.69% of the theophylline molecules are characterized at the wavelengths 331, 587 and 815 nm respectively at the given effective sample sizes. For tablets consisting of the LT formulation, 0.61, 1.05 and 0.54% of the theophylline molecules are characterized. These calculations were performed assuming a homogeneous API distribution within the tablet, which is verified in the following. Based on the experimental setup and the measurement constraints of the system, the actual effective sample size is generally assumed to be underestimated.

Table 3: Effective sample sizes evaluated using the information depths for 331, 587 and 815 nm.

| Formulation | effective sample size $V_{effective}(y(R_\infty))$ mm$^3$ | | |
|---|---|---|---|
| | 331 nm | 587 nm | 815 nm |
| MCC | 1.40 | 0.95 | 1.95 |
| MCC coarse | 1.51 | 0.95 | 2.01 |
| MCC fine | 0.89 | / | 1.73 |
| MT | 1.00 | 0.84 | 1.39 |
| Lac | 1.45 | 1.56 | 1.06 |
| LT | 1.23 | 1.87 | 1.09 |

### 3.7. Spatial API distribution

In order to demonstrate that the measurement of tablet CQAs by UV/Vis diffuse reflectance spectroscopy is representative despite the small sample size, an analysis of the spatial distribution of the API within the LT tablet formulation was performed. The results of this investigation are presented in the following. In Figure 8 a and b, the segmentation results obtained from the U-Net are presented in both 2D and 3D. The segmentation appears visually promising, as the API regions are clearly highlighted. The segmentation was validated against additionally hand-labeled validation data by means of the Dice similarity coefficient DSC [36,37] given by

$$DSC(X, Y) = \frac{2|X \cap Y|}{|X|+|Y|}, \quad (10)$$

where |A| denotes the number of pixels of a pixelized data set A in R$^2$. Here, X and Y represent the API phases within the hand-labeled region in the network output and the validation data, respectively. The Dice similarity coefficient is 0.96, indicating a high level of agreement between the predicted and reference segmentations.

A visual inspection of Figure 8 c does not reveal any trends in the spatial distribution of the API. Figure 9 provides a more quantitative evaluation of this observation. The 2D probability density plot shows the distribution of the local fraction of the API along the tablet's depth (z-axis). The local fraction of the API is defined as the volume fraction of the API within a cube of side length 407.5 μm. This quantity is computed by partitioning the whole 3D image into space filling, non-overlapping cubes that fit completely into the volume of the cylindrical sample. Figure 9 shows that many subvolumes contain a low API fraction of less than 4%, as indicated by the dark blue color, while others show a local volume fraction of more than 10%. Note that the total volume fraction of the segmented API is slightly lower than the volume fraction of the API in the powder formulation. However, a change in volume fraction after compaction and segmentation may occur. Since we are only interested in the analysis of heterogeneity along the tablet, this does not influence the results. In Figure 9 it can be observed that the distribution and the heterogeneity of the API remain consistent along the z-axis of the measured volume, suggesting spatial independence. This observation provides strong evidence that the results obtained via UV/Vis analysis are reliable for in-line product quality quantification, even when it only can assess the tablet's composition near the tablets surface.

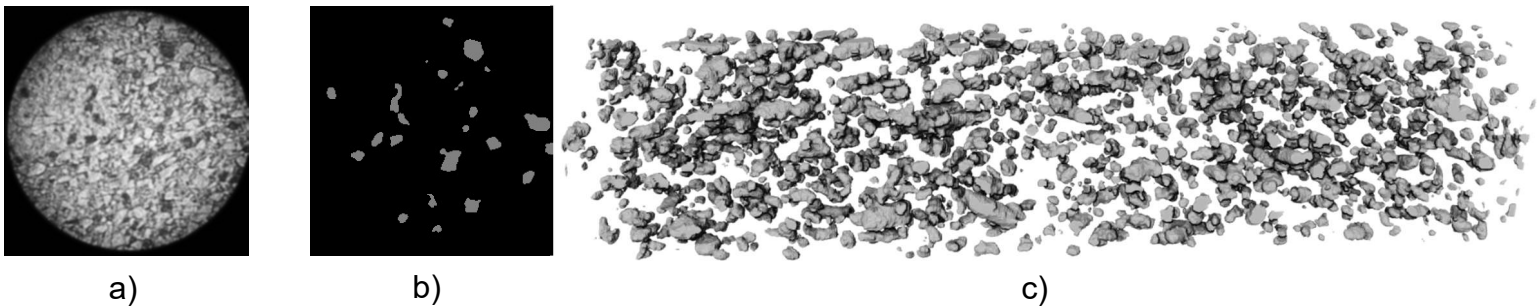

a)   b)   c)

Figure 8: 2D slice of preprocessed CT-image data (a), corresponding segmentation of the neural network (b), and 3D visualization of segmented API (c).

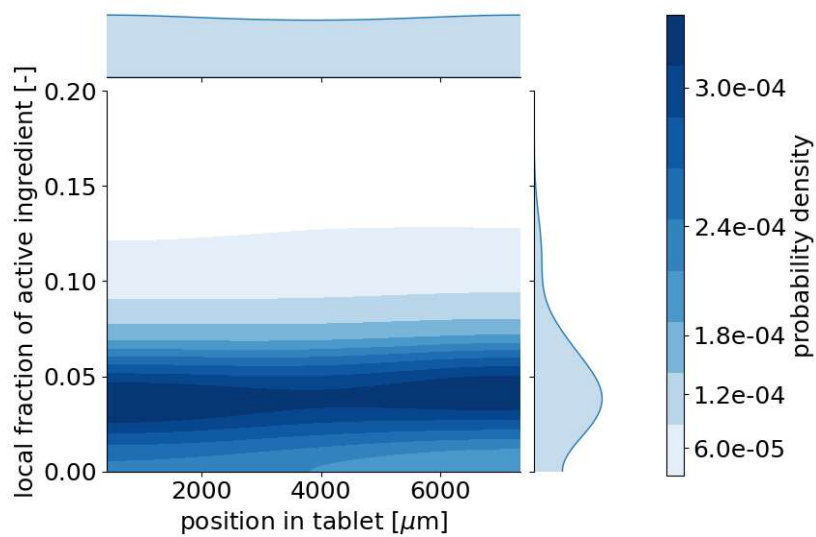

Figure 9: 2D probability density plot of the API distribution along the z-axis of the LT tablet formulation. The plot visualizes the joint distribution of subvolume positions and their API fractions using a 2D kernel density estimation (KDE). The 1D KDE plots (top and right) represent the marginal distributions of the two variables. The absence of significant variation along the z-axis suggests that the two marginal distributions are independent.

## 4. Conclusion

In general, the main objective of this study was to characterize the penetration depth and the effective sample size of UV/Vis radiation in pharmaceutical tablets. Therefore, bilayer tablets consisting of different formulations were produced with different upper layer thicknesses. Here, the lower layer was kept constant in composition and thickness. The investigated bilayer tablets exhibit experimental penetration depths between 0.1 and 0.4 mm. Additionally, the Kubelka-Munk model was utilized to determine the theoretical penetration depths reaching values from 0.25 to 1.38 mm.

In general, a wavelength dependency was found, while differences were observed in the UV, Vis and low NIR range. Thereby, an influence of the composition, the deformation behavior and the particle size was identified. In general, smaller particle sizes leads to a reduction in penetration depth. The effective sample sizes were calculated based on the penetration depth data.

An analysis of the spatial distribution of the API within the LT tablet formulation was performed to demonstrate that the measurement of the API content by UV/Vis diffuse reflectance spectroscopy was representative despite the small sample size using micro-CT. Thereby, a uniform distribution of the API across the tablet was found. Consequently, the penetration depth found and the associated effective sample size do not represent a limitation for UV/Vis spectroscopy when applied as a real-time release tool.

In conclusion, UV/Vis spectroscopy is a promising alternative for real-time monitoring in tableting.

## 5. Acknowledgments

The authors acknowledge the German Research Foundation for funding this project.

## 6. Funding

The authors gratefully acknowledge funding by the German Research Foundation (DFG) within SPP 2364 under grants TH 1817/8-1, PE 1160/37-1 and SCHM 997/47-1.

## 7. Data Availability

The micro-CT data (raw and segmented) that support the findings of this study are openly available via the Open Access Repository and Archive for Research Data of Saxon Universities (OPARA) at: https://doi.org/10.25532/OPARA-797.

**References**
1. Hole, G.; Hole, A.S.; McFalone-Shaw, I. Digitalization in pharmaceutical industry: What to focus on under the digital implementation process? *International Journal of Pharmaceutics: X* **2021**, *3*, 100095, doi:10.1016/j.ijpx.2021.100095.

2. Chen, Y.; Yang, O.; Sampat, C.; Bhalode, P.; Ramachandran, R.; Ierapetritou, M. Digital twins in pharmaceutical and biopharmaceutical manufacturing: A literature review. *Processes* **2020**, *8*, 1088, doi:10.3390/pr8091088.
3. Arden, N.S.; Fisher, A.C.; Tyner, K.; Yu, L.X.; Lee, S.L.; Kopcha, M. Industry 4.0 for pharmaceutical manufacturing: Preparing for the smart factories of the future. *International Journal of Pharmaceutics* **2021**, *602*, 120554, doi:10.1016/j.ijpharm.2021.120554.
4. FDA. *Guidance for Industry PAT - A Framework for Innovative Pharmaceutical Development, Manufacturing, and Quality Assurance*, 2004.
5. Markl, D.; Warman, M.; Dumarey, M.; Bergman, E.-L.; Folestad, S.; Shi, Z.; Manley, L.F.; Goodwin, D.J.; Zeitler, J.A. Review of real-time release testing of pharmaceutical tablets: State-of-the art, challenges and future perspective. *International Journal of Pharmaceutics* **2020**, *582*, 119353, doi:10.1016/j.ijpharm.2020.119353.
6. ICH. *Q8 (R2) Pharmaceutical development;* ICH-Homepage, 2009.
7. ICH. *Q10 Pharmaceutical Quality System;* ICH-Homepage, 2008.
8. Ervasti, T.; Simonaho, S.-P.; Ketolainen, J.; Forsberg, P.; Fransson, M.; Wikström, H.; Folestad, S.; Lakio, S.; Tajarobi, P.; Abrahmsén-Alami, S. Continuous manufacturing of extended release tablets via powder mixing and direct compression. *International Journal of Pharmaceutics* **2015**, *495*, 290–301, doi:10.1016/j.ijpharm.2015.08.077.
9. Bakeev, K.A. *Process Analytical Technology*: *Spectroscopic Tools and Implemented Strategies for the Chemical and Pharmaceutical Industries / edited by Katherine A. Bakeev*, 2nd ed.; Wiley-Blackwell: Oxford, 2010, ISBN 978-0-470-72207-7.
10. Zhong, L.; Gao, L.; Li, L.; Zang, H. Trends-process analytical technology in solid oral dosage manufacturing. *Eur. J. Pharm. Biopharm.* **2020**, *153*, 187–199, doi:10.1016/j.ejpb.2020.06.008.
11. Beer, T. de; Burggraeve, A.; Fonteyne, M.; Saerens, L.; Remon, J.P.; Vervaet, C. Near infrared and Raman spectroscopy for the in-process monitoring of pharmaceutical production processes. *International Journal of Pharmaceutics* **2011**, *417*, 32–47, doi:10.1016/j.ijpharm.2010.12.012.
12. Pauli, V.; Roggo, Y.; Pellegatti, L.; Nguyen Trung, N.Q.; Elbaz, F.; Ensslin, S.; Kleinebudde, P.; Krumme, M. Process analytical technology for continuous manufacturing tableting processing: A case study. *Journal of Pharmaceutical and Biomedical Analysis* **2019**, *162*, 101–111, doi:10.1016/j.jpba.2018.09.016.
13. Järvinen, K.; Hoehe, W.; Järvinen, M.; Poutiainen, S.; Juuti, M.; Borchert, S. In-line monitoring of the drug content of powder mixtures and tablets by near-infrared spectroscopy during the continuous direct compression tableting process. *European Journal of Pharmaceutical Sciences* **2013**, *48*, 680–688, doi:10.1016/j.ejps.2012.12.032.
14. Shah, R.B.; Tawakkul, M.A.; Khan, M.A. Process analytical technology: chemometric analysis of Raman and near infra-red spectroscopic data for predicting physical properties of extended release matrix tablets. *Journal of Pharmaceutical Sciences* **2007**, *96*, 1356–1365, doi:10.1002/jps.20931.
15. Peeters, E.; Da Tavares Silva, A.F.; Toiviainen, M.; van Renterghem, J.; Vercruysse, J.; Juuti, M.; Lopes, J.A.; Beer, T. de; Vervaet, C.; Remon, J.-P. Assessment and prediction of tablet properties using transmission and backscattering Raman spectroscopy and transmission NIR spectroscopy. *Asian Journal of Pharmaceutical Sciences* **2016**, *11*, 547–558, doi:10.1016/j.ajps.2016.04.004.
16. Donoso, M.; Kildsig, D.O.; Ghaly, E.S. Prediction of tablet hardness and porosity using near-infrared diffuse reflectance spectroscopy as a nondestructive method. *Pharmaceutical Development and Technology* **2003**, *8*, 357–366, doi:10.1081/PDT-120024689.
17. Man, A. de; Uyttersprot, J.-S.; Chavez, P.-F.; Vandenbroucke, F.; Bovart, F.; Beer, T. de. The application of near-infrared spatially resolved spectroscopy in scope of achieving continuous real-time quality monitoring and control of tablets with challenging dimensions. *International Journal of Pharmaceutics* **2023**, *641*, 123064, doi:10.1016/j.ijpharm.2023.123064.
18. Clarke, F.C.; Hammond, S.V.; Jee, R.D.; Moffat, A.C. Determination of the information depth and sample size for the analysis of pharmaceutical materials using reflectance near-infrared microscopy. *Applied Spectroscopy* **2002**, *56*, 1475–1483, doi:10.1366/00037020260377797.
19. Laborde, A.; Jaillais, B.; Bendoula, R.; Roger, J.-M.; Jouan-Rimbaud Bouveresse, D.; Eveleigh, L.; Bertrand, D.; Boulanger, A.; Cordella, C.B.Y. A partial least squares-based approach to assess the light penetration depth in wheat flour by near infrared hyperspectral imaging. *Journal of Near Infrared Spectroscopy* **2020**, *28*, 25–36, doi:10.1177/0967033519891594.
20. Huang, M.; Kim, M.S.; Chao, K.; Qin, J.; Mo, C.; Esquerre, C.; Delwiche, S.; Zhu, Q. Penetration depth measurement of near-infrared hyperspectral imaging light for milk powder. *Sensors* **2016**, *16*, 441, doi:10.3390/s16040441.
21. Iyer, M.; Morris, H.R.; Drennen, J.K. Solid Dosage Form analysis by near infrared spectroscopy: Comparison of reflectance and transmittance measurements including the determination of effective sample mass. *Journal of Near Infrared Spectroscopy* **2002**, *10*, 233–245, doi:10.1255/jnirs.340.
22. Saeed, M.; Saner, S.; Oelichmann, J.; Keller, H.; Betz, G. Assessment of diffuse transmission mode in near-infrared quantification--part I: The press effect on low-dose pharmaceutical tablets. *Journal of Pharmaceutical Sciences* **2009**, *98*, 4877–4886, doi:10.1002/jps.21777.
23. Saeed, M.; Probst, L.; Betz, G. Assessment of diffuse transmission and reflection modes in near-infrared quantification, part 2: DIFFuse reflection information depth. *Journal of Pharmaceutical Sciences* **2011**, *100*, 1130–1141, doi:10.1002/jps.22344.
24. Schlindwein, W.; Bezerra, M.; Almeida, J.; Berghaus, A.; Owen, M.; Muirhead, G. In-line UV-Vis Spectroscopy as a Fast-Working Process Analytical Technology (PAT) during Early Phase Product development using hot melt extrusion (HME). *Pharmaceutics* **2018**, *10*, 166, doi:10.3390/pharmaceutics10040166.
25. Brands, R.; Tebart, N.; Thommes, M.; Bartsch, J. UV/Vis spectroscopy as an in-line monitoring tool for tablet content uniformity. Journal of Pharmaceutical and Biomedical Analysis **2023**, *236*, 115721, doi:10.1016/j.jpba.2023.115721.
26. Brands, R.; Le, T.N.; Bartsch, J.; Thommes, M. A Step Towards real-time release testing of pharmaceutical tablets: Utilization of CIELAB color space. *Pharmaceutics* **2025**, *17*, 311, doi:10.3390/pharmaceutics17030311.
27. KUBELKA, P. New contributions to the optics of intensely light-scattering materials. *Journal of the Optical Society of America* **1948**, *38*, 448–457, doi:10.1364/JOSA.38.000448.
28. Emmel, P. Nouvelle formulation du modèle de Kubelka et Munk avec application aux encres fluorescentes. *Ecole de Printemps 2000 - Le Pays d'Apt en Couleurs* **2000**, 87–96.
29. *Europäisches Arzneibuch 10. Ausgabe;* Deutscher Apotheker Verlag, 2020.
30. Barnes, R.J.; Dhanoa, M.S.; Lister, S.J. Standard normal variate transformation and de-trending of near-infrared diffuse reflectance spectra. *Applied Spectroscopy* **1989**, *43*, 772–777, doi:10.1366/0003702894202201.
31. La Alcaraz de Osa, R.; Iparragirre, I.; Ortiz, D.; Saiz, J.M. The extended Kubelka–Munk theory and its application to spectroscopy. *ChemTexts* **2020**, *6*, 1–14, doi:10.1007/s40828-019-0097-0.
32. Çiçek, Ö.; Abdulkadir, A.; Lienkamp, S.S.; Brox, T.; Ronneberger, O. 3D U-Net: Learning dense volumetric segmentation from sparse annotation. In . International

Conference on Medical Image Computing and Computer-Assisted Intervention; Springer, Cham, **2016**; 424–432, ISBN 978-3-319-46723-8.
33. Buades, A.; Coll, B.; Morel, J.-M. A non-local algorithm for image denoising, In 2005 IEEE Computer Society Conference on Computer Vision and Pattern Recognition **2005** 60–65, doi:10.1109/CVPR.2005.38.
34. Lin, T.-Y.; Goyal, P.; Girshick, R.; He, K.; Dollar, P. Focal Loss for Dense Object Detection, In *Proceedings of the IEEE international conference on computer vision* 2017 2999–3007, doi:10.1109/ICCV.2017.324.
35. Furat, O.; Kirstein, T.; Leißner, T.; Bachmann, K.; Gutzmer, J.; Peuker, U.A.; Schmidt, V. Multidimensional characterization of particle morphology and mineralogical composition using CT data and R-vine copulas. *Minerals Engineering* **2024**, *206*, 108520, doi:10.1016/j.mineng.2023.108520.
36. Dice, L.R. Measures of the Amount of Ecologic Association Between Species. *Ecology* **1945**, *26*, 297–302, doi:10.2307/1932409.
37. Carass, A.; Roy, S.; Gherman, A.; Reinhold, J.C.; Jesson, A.; Arbel, T.; Maier, O.; Handels, H.; Ghafoorian, M.; Platel, B.; et al. Evaluating white matter lesion segmentations with refined Sørensen-Dice analysis. *Scientific reports* **2020**, *10*, 8242, doi:10.1038/s41598-020-64803-w.